\documentstyle[12pt,epsfig]{article}
\def\b{\bar}
\def\d{\partial}

\def\cA{{\cal A}}
\def\cD{{\cal D}}

\def\m{\mu}
\def\n{\nu}

\def\t{\tau}

\def\~{\tilde}
\def\h{\eta}

\def\bY3{\bar Y_{,3}}
\def\Y3{Y_{,3}}
\def\z{\zeta}
\def\Z{{\b\zeta}}
\def\Y{{\bar Y}}
\def\cZ{{\bar Z}}
\def\`{\dot}
\def\be{\begin{equation}}
\def\ee{\end{equation}}
\def\bea{\begin{eqnarray}}
\def\eea{\end{eqnarray}}

\def\fn{\footnote}

\def\cF{{\cal F}}

\def\mn{{\mu\nu}}

\begin{document}
%\twocolumn

\title{Two Stringy Systems of the Kerr Spinning Particle}

\author{Alexander Burinskii\\
Gravity Research Group, NSI Russian
Academy of Sciences\\
B. Tulskaya 52, 115191 Moscow, Russia}
\maketitle

\begin{abstract}
\noindent
A classical spinning particle based on the
Kerr-Newman black hole (BH) solution  is considered.
For  parameters of spinning particles $|a|>>m$, the BH horizons
disappear and BH image is drastically changed. We show that it turns
into a skeleton formed by two coupled stringy systems.
One of them is the Kerr singular ring which
can be considered as a circular D-string with an orientifold
world-sheet.
Analyzing the aligned to the Kerr congruence electromagnetic
excitations of this string, we obtain the second stringy system
which consists of two axial half-infinite chiral D-strings.
These axial strings are similar to the Dirac monopole strings
but carry the induced chiral traveling pp-waves.
Their field structure can be described by the field model suggested
by Witten for the cosmic superconducting strings.
We discuss a relation of this stringy system to the Dirac equation
 and argue that this stringy system can play a role of a
classical carrier of the wave function.

\end{abstract}

\section{Introduction}

The Kerr rotating black hole solution displays some remarkable
relations to spinning particles
\cite{Car,Isr,Bur0,IvBur,Lop,BurSen,BurStr,BurSup,BurBag}. For the
parameters of elementary particles, $|a|>> m$, and black-hole
horizons disappear. This  changes drastically the usual black hole
image since there appear  the rotating source
in the form of a closed singular ring of the Compton radius $ a=J/m$.
\fn{Here $J$ is angular momentum and $m$ is mass. We use the units $c=\hbar
=G=1$, and signature $(-+++)$.}.  In the model of the Kerr
spinning particle - ``microgeon" \cite{Bur0} this ring was considered as a
gravitational waveguide  leading the traveling electromagnetic (and
fermionic) wave excitations. The assumption that the Kerr singular
ring represents a closed relativistic string was advanced about
thirty years ago \cite{IvBur}, which got confirmation on the level
of the evidence of Refs. \cite{BurSen,IvBur1,Bur1}. However, the attempts
to show it explicitly ran into obstacles which were related with
the very specific motion of the Kerr ring - the lightlike sliding
along itself.  It could be described as a string containing
lightlike modes of only one direction.  However, the relativistic
string equations do not admit such solutions.

In previous paper \cite{BurOri} we resolved this problem showing
that the Kerr ring satisfies all the stringy equations representing
a string with an orientifold structure.

In this paper we consider consequences of the electromagnetic excitations
of the Kerr circular string and
find out an unavoidable appearance of one or two axial half-infinite strings
which are topologically coupled to the Kerr ring and similar to
the Dirac monopole string. These strings  carry the chiral
traveling waves induced by the e.m. excitations of the Kerr circular string.

\begin{figure}[ht]
\centerline{\epsfig{figure=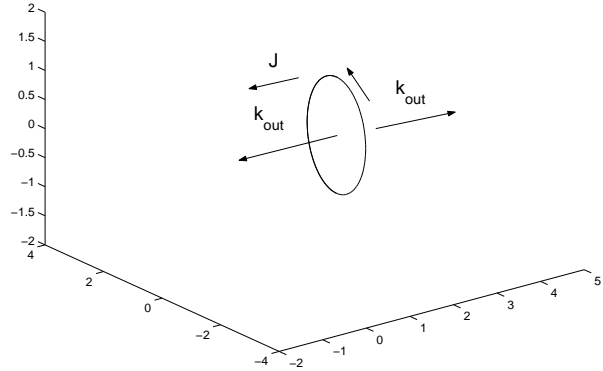,height=6cm,width=8cm}}
\caption{Stringy skeleton of the Kerr spinning particle.
Circular D-string and the directed outwards two axial half-infinite chiral
D-strings.}
\end{figure}

Indeed, the frame of the Kerr spinning particle
consists of two topologically coupled stringy systems. The appearance of the
axial half-infinite strings looks strange at first sight. Meanwhile, we
obtain that it can be a new and very important element of
the structure of spinning particles.
 For the moving particle the excitations of the chiral strings are
modulated by de Broglie periodicity and therefore, the axial strings turn out
to be the carriers of de Broglie wave.

In the zone which is close to the Kerr string, our treatment is based on
the Kerr-Schild formalism  \cite{DKS}
and previous paper \cite{Bur-nst} where the real and complex
structures of the Kerr geometry were considered. For the reader convenience
 we describe briefly the necessary details of these structures.
Meanwhile, in the far zone, structure of this string is
described by the very simple class of pp-wave solutions \cite{KraSte,Per}.
The resulting stringy frame turns to be very simple and easy
for description. We obtain that these strings belongs to the class of
the chiral superconducting strings
which have recently paid considerable attention in astrophysics.

\section{The structure of the Kerr congruence. Microgeon.}

We use the Kerr-Schild approach to the Kerr geometry \cite{DKS},
which is based on the Kerr-Schild form of the metric \be g_{\m\n}
= \h_{\m\n} + 2 h k_{\m} k_{\n}, \label{ksa} \ee where $ \h_\mn $
is the metric of auxiliary Minkowski space-time, $ h= \frac
{mr-e^2/2} {r^2 + a^2 \cos^2 \theta},$ and $k_\m$ is a twisting
null field, which is tangent to the Kerr principal null congruence
(PNC) and is determined by the form \fn{The rays of the Kerr PNC
are twistors and the Kerr PNC is determined by the Kerr theorem as
a quadric in projective twistor space \cite{Bur-nst}.}
 \be k_\m dx^\m = dt +\frac z r dz + \frac r {r^2 +a^2} (xdx+ydy) - \frac a
{r^2 +a^2} (xdy-ydx) .  \label{km} \ee The form of the Kerr PNC is
shown in Fig. 1.  It follows from Eq.(\ref{ksa}) that the field
$k^\m$ is null with respect to $\h_\mn$ as well as with respect to
the full metric $g_\mn$, \be k^\m k_\m = k^\m k^\n g_\mn =  k^\m
k^\n \eta_\mn. \label{kgh} \ee

\begin{figure}[ht]
\centerline{\epsfig{figure=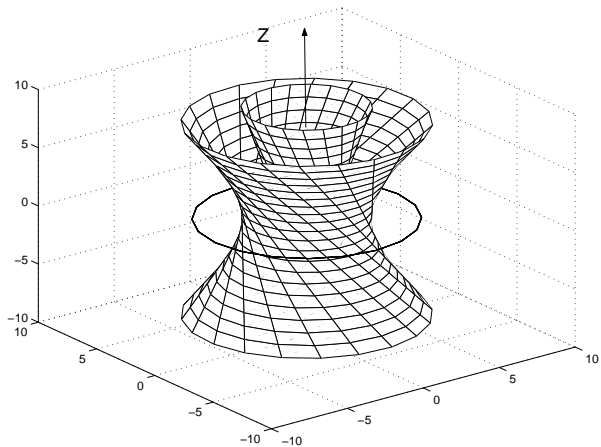,height=6cm,width=8cm}}
\caption{The Kerr singular ring and 3D section of the Kerr
principal null congruence. Singular ring is a branch line of
space, and PNC propagates from the ``negative'' sheet of the Kerr
space to the ``positive '' one, covering the space-time twice. }
\end{figure}

The metric is singular at the ring $r=\cos\theta=0$, which is
the focal region of the oblate spheroidal coordinate system
$r, \theta, \phi$.

  The Kerr singular ring is the branch line of the
Kerr space on two folds: positive sheet ($r>0$) and `negative' one
($r<0$). Since for $|a|>>m$ the horizons disappear, there appears
the problem of the source of the Kerr solution with the
alternative: either to remove this twofoldedness or to give it a
physical interpretation.  Both approaches have received attention,
and it seems that both are valid for different models. The most
popular approach was connected with the truncation of the negative
sheet of the Kerr space, which leads to the source in the form of
a relativistically rotating disk \cite{Isr} and to the class of
the disklike \cite{Lop} or baglike \cite{BurBag} models of the Kerr
spinning particle.

 An alternative way is to retain the negative sheet treating it as
 the sheet of advanced fields. In this case the source of the spinning
particle turns out to be the Kerr singular ring with the
electromagnetic excitations in the form of traveling waves, which
generate spin and mass of the particle.  A model of this sort was
suggested in 1974 as a model of ``microgeon with spin"
\cite{Bur0}. The Kerr singular ring was considered as a
 waveguide providing a circular propagation of an electromagnetic or
fermionic wave excitation. Twofoldedness of the Kerr geometry
admits the integer and half integer excitations  with $n=2\pi
a/\lambda$ wave periods on the Kerr ring of radius $a$, which
turns out to be consistent with the corresponding values of the
Kerr parameters $m= J/a$.

The lightlike structure of the Kerr ring worldsheet is seen from
the analysis of the Kerr null congruence near the ring. The
lightlike rays of the Kerr PNC are tangent to the ring.

It was recognized long ago \cite{IvBur} that the Kerr singular ring
 can be considered in the Kerr spinning particle as a string with
traveling waves.  One of the most convincing evidences obtained
by the analysis of the axidilatonic generalization of the Kerr
solution (given by Sen \cite{Sen}) near the Kerr singular ring
was given in \cite{BurSen}. It was shown that the fields
 near the Kerr ring are very similar to the field around a
heterotic string.

\section{The Kerr Orientifold Worldsheet}

One can see that the worldsheet of the Kerr ring satisfies the
bosonic string equations and constraints; however, there appear
problems with boundary conditions.
In this section we recall briefly the analysis given in \cite{BurOri}.

 The general solution of the string wave equation
$(\frac {\d ^2} {\d \sigma ^2} - \frac {\d ^2} {\d \t ^2}) X^\m  =0$
can
be represented as the sum of the `left' and `right' modes:
$X^\m(\sigma, \t)=
X_R^\m (\t -\sigma) + X_L^\m (\t +\sigma), $ and the oscillator expansion is
\be X_R^\m (\t -\sigma) =\frac 12 [ x^\m + l^2 p^\m (\t -\sigma) +
il\sum_{n\ne0} \frac 1n \alpha_n^\m e^{-2in(\t-\sigma)}] ,
\label{R} \ee

\be X_L^\m (\t +\sigma) =\frac 12 [ x^\m + l^2 p^\m (\t +\sigma) +
il\sum _{n\ne0} \frac 1n \tilde\alpha_n^\m e^{-2in(\t+\sigma)}] ,
\label{L} \ee

where $ l =\sqrt{2\alpha^\prime} =\frac 1{\sqrt{\pi T}}$,  $T$ is tension,
$x^\m$ is position of center of mass, and $p^\m$ is momentum of string.

The string constraints $\dot X_\m \dot X^\m + X^{\prime}_\m
X^{\prime\m} =0, \qquad \dot X_\m  X^{\prime\m} =0,$ are satisfied
if the modes are lightlike [$()^\prime \equiv \d _\sigma ()$], \be
(\d _ \sigma X_{L(R)\m}) (\d _{\sigma} X_{L(R)}^\m) =0. \label{c1}
\ee Setting $2\sigma = a \phi$ one can describe the lightlike
worldsheet of the Kerr ring (in the rest frame of the Kerr
particle) by the surface \be X_L^\m(t,\sigma) = x^\m + \frac 1{\pi
T} \delta _0^\m p^0 (t + \sigma) + \frac a 2 [(m^\m +in^\m)
e^{-i2(\t+\sigma)} + (m^\m -in^\m) e^{i2(\t+\sigma)}]  ,
\label{kring} \ee where $m^\m$ and $n^\m$ are two spacelike basis
vectors lying in the plane of the Kerr ring.  One can see that \be
X_L^{\prime\m} = \frac 1{\pi T} \delta _0^\m  p^0  + 2a[-m^\m \sin
2(\t+\sigma) + n^\m \cos 2(\t+\sigma)] \label{dkring} \ee will be
a light-like vector if one sets $p^0 = 2\pi a T $. It shows that
the Kerr worldsheet could be described by modes of one (say
``left") null direction. The solution $X(\t, \sigma) =X_L(\t+
\sigma)$
 satisfies the string wave equation and constraints, but
there appears the problem with boundary conditions.
The closed string boundary condition
\be X^\m(\t,\sigma)=X^\m(\t,\sigma+\pi)
\label{cl-b} \ee
 will not be satisfied since the time component
$X_L^0 (t,\sigma +\pi)$ acquires contribution from the second term
in (\ref{kring}), which is usually compensated by this term from
the `right' mode. The familiar boundary conditions for the open
strings follow from the condition of cancelling of the surface term
$ -T\int d\t [X^\prime _\m \delta X^\m |_{\sigma=\pi} - X^\prime
_\m \delta X^\m |_{\sigma=0} ]$ in the string action  \cite{GSW},
and are \be X^{\prime \m}(\t,0)= X^{\prime \m}(\t, \pi)=0,
\label{op-b} \ee
which also demand both types of modes to form
standing waves. However, this demand can be weakened  to
\be
X^{\prime \m}(\t,0)= X^{\prime \m}(\t ,\pi). \label{Kop-b} \ee

It seems that the lightlike oriented string can contain traveling
waves of only one direction if we assume that it is open, but has
the joined ends. However, the ends  $\sigma=0,$ and $\sigma=\pi$
are not joined  indeed.

These difficulties can be removed by the formation of the
worldsheet orientifold.

It is well known \cite{GSW} that the interval of an open string  $\sigma
\in [0,\pi] $ can be formally extended to $[0,2\pi]$, setting
\be X_R (\sigma
+\pi) =X_L (\sigma), \qquad X_L (\sigma +\pi) =X_R (\sigma).  \label{ext} \ee

By such an extension, the both types of modes, ``right" and
``left", will appear in our case since the ``left" modes will play
the role of ``right" ones on the extended piece of interval. If
the extension is completed by the changing of orientation on the
extended piece,  $\sigma ^\prime = \pi - \sigma $, with a
subsequent identification of $\sigma$ and  $\sigma ^\prime$, then
one obtains the closed string on the interval $[0,2\pi]$ which is
 folded and takes the form of the initial open string.

Formally, the worldsheet orientifold represents a doubling of the
worldsheet with the orientation reversal on the second sheet. The
fundamental domain $[0,\pi]$ is extended to $\Sigma=[0,2\pi]$ with
formation of folds at the ends of the interval $[0,\pi]$.

\begin{figure}[ht]
\centerline{\epsfig{figure=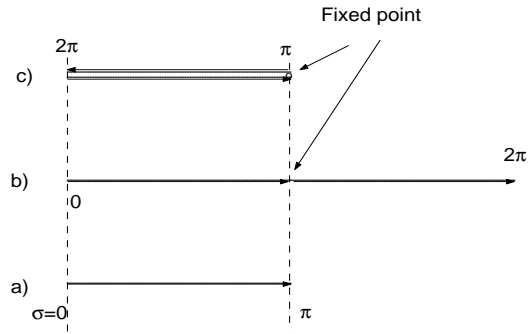,height=5cm,width=7cm}}
\caption{Formation of orientifold: a) the initial string interval;
b) extension of the interval and formation of the both side
movers; c) formation of the orientifold.} \end{figure}

\section{Solution of the e.m. field equations}
\bigskip
To realize the idea of the Kerr spinning particle as a ``microgeon''
we have to consider the  electromagnetic excitations of the Kerr string
which are described by the solutions which are aligned to the Kerr PNC
on the Kerr background.

The treatment on this section is based on the Kerr-Schild formalism,
and the readers which are not aware of this formalism can omit this part
by first reading going to the physical consequences of these solutions.

The aligned field equations for the Einstein-Maxwell system in the
Kerr-Schild class were obtained in \cite{DKS}.
Electromagnetic field is given by tetrad components of self-dual tensor
\be \cF _{12} =AZ^2 \label{1}\ee
\be \cF _{31} =\gamma Z - (AZ),_1  \ . \label{2}\ee
The equations for electromagnetic field are
\be A,_2 - 2 Z^{-1} \cZ Y,_3 A  = 0 , \label{3}\ee
\be \cD A+  \cZ ^{-1} \gamma ,_2 - Z^{-1} Y,_3 \gamma =0 .
\label{4}\ee
Gravitational field equations yield
\be M,_2 - 3 Z^{-1} \cZ Y,_3 M  = A\bar\gamma \cZ ,  \label{5}\ee
\be \cD M  = \frac 12 \gamma\bar\gamma  , \label{6}\ee
where
\be
\cD=\d _3 - Z^{-1} Y,_3 \d_1 - \cZ ^{-1} \Y ,_3 \d_2   \ . \label{cD}
\ee
Solutions of this system were given in \cite{DKS} only for stationary case
for $\gamma=0$.
Here we consider the oscillating electromagnetic
solutions which corresponds to the case $\gamma \ne 0$.

For the sake of simplicity we have to consider the gravitational
Kerr-Schild field as stationary, although in the resulting e.m.
solutions the axial symmetry is broken, which has to lead to
oscillating backgrounds if the back reaction is taken into account.

The recent progress in the obtaining the nonstationary solutions of the
Kerr-Schild class is connected with introduction of a complex retarded time
parameter $\t = t_0 +i\sigma = \t |_L$ which is determined as a result of the
intersection of the left (L) null plane
and the complex world line \cite{Bur-nst}. The left null planes are the
left generators of the complex null cones and play a role of the null cones
in the complex retarded-time construction.
The $\t$ parameter satisfies to the relations
\be
(\t),_2 =(\t),_4 = 0 \ . \label{7}
\ee
It allows one to represent the equation (\ref{3}) in the form
\be
(AP^2),_2=0 \ , \label{8}
\ee
and to get the following general solution
\be A= \psi(Y,\t)/P^2
\label{A}\ee
which has the form obtained in \cite{DKS}.
The only difference is in the extra dependence of the
function $\psi$ from the retarded-time parameter $\t$.

It was shown in \cite{Bur-nst} that action of operator $\cD$ on the variables
$Y, \bar Y $ and $ \rho$ is following
\be \cD Y = \cD \bar Y = 0,\qquad
\cD \rho =1 \ , \label{10}\ee
and therefore
$\cD \rho = \d \rho / \d t_0 \cD t_0  = P\cD t_0 =1 $,
that yields
\be
\cD t_0 = P^{-1} .
\ee
As a result the equation (\ref{4}) takes the form
\be
\dot A = -(\gamma P),_{\bar Y} , \label{11}
\ee
where $\dot {( \ )} \equiv \d_{t_0}$.

For considered here stationary background $P=2^{-1/2}(1+Y\bar Y)$, and
$\dot P = 0$.  The coordinates $Y$,  and $\t$ are independent from
$\bar Y$, which allows us to integrate Eq. (\ref{11}) and
we obtain the following general solution
\be
\gamma  = - P^{-1}\int \dot A d\bar Y =
 - P^{-1}\dot \psi (Y,\t) \int  P^{-2}d\bar Y  =
 \frac{2^{1/2}\dot \psi} {P^2 Y} +\phi (Y,\t)/P ,
\label{12}\ee
where $\phi$ is an arbitrary analytic function of $Y$ and $\t$.

The term $\gamma$  in
$ \cF _{31} =\gamma Z - (AZ),_1  \ $
describes a part of the null
electromagnetic radiation   which
falls of asymptotically as $1/r$ and
propagates along the Kerr principal null congruence $e^3$.
As it was discussed in \cite{Bur-nst,BurAli} it describes
a loss of mass by radiation with the stress-energy tensor
$\kappa T^{(\gamma)}_\mn = \frac 12 \gamma \bar \gamma e^3_{\m} e^3_{\n}$
and has to lead to an infrared divergence.
However,  the Kerr twofoldedness
and the structure of the Kerr principal null congruence show us that the
loss of mass on the positive sheet of metric is really compensated by an
opposite process on the ``negative" sheet of the Kerr space where is an
in-flow of the radiation. In the microgeon model
\cite{Bur-nst,BurOri,BurAli}, this field acquires interpretation of the vacuum zero
point field $T^{(\gamma)}_\mn = <0|T_{\mn}|0>$.
Similar to the treatment of the zero point field in the
Casimir effect one has to regularize stress energy tensor
by the subtraction
\be T^{(reg)}_{\mn} = T_{\mn} - <0|T_{\mn}|0>, \ee
under the condition $ T^{(\gamma) \ \mn} ,_\m = 0$ which is
satisfied for the $\gamma$ term.

Let's now consider in details the second term in (\ref{2}):
\be
(AZ),_1 = (Z/P)^2 (\psi ,_Y - 2 \psi P_Y) +
(Z/P^2) \dot \psi \t,_1 + A Z,_1 .
\ee
For stationary case we have relations $Z,_1 =2ia \bar Y (Z/P)^3 $
and  $\t,_1 =- 2ia \bar Y Z/P^2 $ (see Appendix).
This yields
\be
(AZ),_1 = (Z/P)^2 (\psi ,_Y - 2ia  \dot \psi \bar Y /P^2 - 2 \psi P_Y/P) +
A 2ia \bar Y (Z/P)^3 .
\label{AZ1}
\ee

Since $Z/P =1/(r+ia \cos \theta)$, this expression contains the terms
which are singular at the Kerr ring and fall off like $r^{-2}$ and  $r^{-3}$.
However, it contains also
the factors which depend on coordinate
$Y = e^{i\phi} \tan \frac {\theta} 2 $ and
can be singular at the z-axis.

These singular factors  can be selected in the full
expression for the aligned e.m. fields and as a result there
appear two  half-infinite lines of
singularity,  $z^+$ and $z^-$,
which correspond to $\theta =0$ and $\theta=\pi$ and coincide
with corresponding axial lightlike rays of the Kerr principal null
congruence.  On the ``positive'' sheet of the Kerr background these two
half-rays are directed outward. However, one can see that they are going
from the ``negative" sheet and  appear on the ``positive'' sheet
passing through the Kerr ring (see Fig. 2).

The general solution for the aligned electromagnetic fields has the form

\be
\cF = \cF _{31} \ e^3 \wedge e^1 + \cF _{12} \ ( e^1 \wedge e^2 +
 e^3 \wedge e^4 ). \label{cFal}
\ee
In the null Cartesian coordinates the Kerr-Schild null tetrad  has the form
\fn{In the
paper \cite{DKS}
treatment is given in terms of the ``in" - going congruence $e^3$
(advanced fields).
 Here we need to use the ``out''-going congruence. The simplest way to do it
 retaining the basic relations of the paper \cite{DKS} is to replace
$t\to -t$ in the definition of the null Cartesian coordinates.
Therefore, we use here the notations
\bea
2^{1\over2}\z &=& x+iy ,\qquad 2^{1\over2} \Z = x-iy , \nonumber\\
2^{1\over2}u &=& z - t ,\qquad 2^{1\over2}v = z + t . \label{ncc}
\eea
}

\begin{eqnarray}
e^1 &=& d \zeta - Y dv, \qquad  e^2 = d \bar\zeta -  \bar Y dv, \nonumber \\
e^3 &=&du + \bar Y d \zeta + Y d \bar\zeta - Y \bar Y dv, \nonumber\\
e^4 &=&dv + h e^3\label{KSt}.
\end{eqnarray}

Evaluating the basis two-forms in the Cartesian coordinates we obtain
\be
e^1 \wedge e^2 +  e^3 \wedge e^4 = d\z \wedge d \Z + du \wedge dv +
Y d\Z  \wedge  dv,
\ee
and
\be
e^3 \wedge e^1 = Y \ d\Z \wedge d \z + du \wedge dz -
Y du  \wedge  dv - Y^2 \ d\Z \wedge d v.
\ee

\section{Axial singular waves}

The obtained general solution for the aligned electromagnetic fields
(\ref{cFal}) contains the factors which depend on coordinate
$Y = e^{i\phi} \tan \frac {\theta} 2 $ and
can be singular at the z-axis.

We will now be interested in the wave terms and omit the terms describing the
longitudinal components and the field $\gamma$.

The wave terms
are proportional to the following basis  two-forms

$e^3 \wedge e^1 |_{wave} = du \wedge d \z + Y^2 dv \wedge d \Z$

and

$ e^1 \wedge e^2 + e^3 \wedge e^4 |_{wave} = d \zeta \wedge d \bar\zeta $.

Near the positive half-axis $z^+$,  we have  $Y\to 0$  and
near the negative half-axis $z^-$,  we have  $Y\to \infty$.

Therefore, with the exclusion the $\gamma$ term, the wave terms
 of the e.m. field (\ref{cFal}) take the form
\be \cF |_{wave} =f_R \ d \z \wedge d u  +
f_L \ d \Z \wedge d v ,
\label{cFLR}
\ee
where the factor
\be
f_R = (AZ),_1
\label{fR}
\ee
describes the ``right"  waves propagating along the $z^+$ half-axis,
and the factor
\be
f_L =2Y \psi (Z/P)^2 + Y^2 (AZ),_1
\label{fL}
\ee
describes the ``left"  waves propagating along the  $z^-$ half-axis,
and some of them are singular at z axis.

Besides, since $Z/P=(r+ia \cos \theta)^{-1}$, all
the terms are also singular at the Kerr ring $r=\cos \theta =0$.
Therefore, the singular excitations of the Kerr ring turn out to be
connected with the axial singular waves.

Let us consider the solutions
describing traveling waves along the Kerr ring
\be
\psi _n (Y,\t) = q Y^n \exp {i\omega \t}
\equiv q (\tan \frac \theta 2)^n \exp {i(n\phi + \omega _n \t)}.
\label{psin}
\ee

Near the Kerr ring one has $\psi =\exp {i(n\phi + \omega t)}$, and
$|n|$ corresponds to the number of the wave lengths along the Kerr ring.
The parameter $n$ has to be integer for the smooth and single-valued
solutions, however, as we shell see bellow,
the half-integer $n$ can be interesting too.

Meanwhile, by $Y\to 0$ one approaches to the positive z-axis where
the solutions may be singular too.
Similar, by $Y\to \infty$ one approaches to the negative z-axis, and
some of the solutions turns out to be singular there.

When considering asymptotical properties
of these singularities by $r \to \infty $, we have
 $z=r\cos \theta$, and for the distance $\rho$ from the $z^+$ axis we have
the expression $\rho = z \tan \theta \simeq 2 r |Y| $ by $Y\to 0$.
Therefore, for the asymptotical region near the $z^+$ axis we have to put
$Y = e^{i\phi} \tan {\frac \theta 2} \simeq  e^{i\phi} \frac \rho {2r}$, and
$|Y|\to 0$,
while for the asymptotical region near the $z^-$ axis
$Y = e^{i\phi} \tan {\frac \theta 2 } \simeq  e^{i\phi} \frac {2r} \rho  $,
and $|Y|\to \infty$.

 The parameter $\t=t -r -ia \cos \theta$ takes near the
z-axis the values
\be \t _+ = \t |_{z^+}= t-z-ia,\quad \t _- = \t |_{z^-}
=t+z +ia.
\label{tauz}
\ee

It has also to be noted that for $|n|>1$ the solutions contain the axial
singularities which do not fall of  asymptotically, but are increasing.
Therefore, we shell restrict the treatment by the cases $|n|\le 1$.

The leading wave terms for $|n|\le 1$ are given in the Appendix.

The leading singular wave for $n=1$ is
\be
\cF^-_1=\frac {4q e^{i2\phi+i\omega _{1} \t_- }} {\rho ^2} \ d \Z \wedge dv.
\ee
It propagates to $z=-\infty$ and has the uniform
axial singularity at $z^-$ of order $\rho ^{-2}$.

Meanwhile, the leading singular wave for $n=-1$ is

\be
\cF^+_{-1}= -
\frac {4q e^{-i2\phi+i\omega _{-1} \t_+ }} {\rho ^2} \ d \z \wedge du ,
\ee

and has the similar uniform axial singularity at $z^+$ which
propagates to $z=+\infty$.

The waves with $n=0$ are regular.

In what follows we will show that these singularities form the half-infinite
chiral  strings, in fact superconducting D-strings.
There are several arguments in favor of the system containing a combination
of two strings of opposite chirality, $n=\pm 1$.

First, if the solution contains only one half-infinite string,
like the Dirac monopole string, it turns out to be asymmetric with respect
to the $z^{\pm}$ half-axis, which leads to a nonstationarity via a recoil.

Then, the symmetric stringy solutions exclude the appearance of monopole
charge.

Note also, that the pure chiral strings, containing modes of only
one direction, cannot exist and any chiral string
has to be connected to some object containing an anti-chiral part.
Indeed, the pure chiral excitation depends only  on one of the
parameters $\t _\pm = t \pm \sigma $, and as a result the
world-sheet is degenerated in a world-line. \fn{This  argument
was suggested by G. Alekseev.} This is seen in the models of the cosmic
chiral strings  where the chiral excitations are joined to
some mass \cite{CarPet} or are sitting on some string having modes
of opposite chirality \cite{Vil}.
In our case the partial pp-wave e.m. excitation has the same
chirality as the half-infinite carrier of this excitation
(the axial ray of PNC).
Therefore, the combination of two $n=\pm 1$ excitations looks very natural
and leads to the appearance of a full stringy system with two
half-infinite singular D-strings of opposite chirality,
``left'' and ``right'', as it is shown at the fig.1.
The world-sheet of the system containing from two straight chiral
strings will be given by
\be
x^\m (t,z)=\frac 12 [ (t-z)k_R^{\m} + (t+z)k_L^{\m}],
\label{astr}
\ee
where the lightlike vectors $k^\m$ are constant and normalized.
At the rest frame the timelike components are equal
$k_R^{0}  =k_L^{0}=1$, and the spacelike components are oppositely directed,
$k_R^{a} + k_L^{a}=1, \quad a=1,2,3.$
Therefore,
$\dot x^\m =(1,0,0,0),$ and $ x'^\m =(0,k^a),$
and the Nambu-Goto string action
\be
S=\alpha^{\prime -1}\int\int\sqrt{(\dot x)^2 ( x')^2 - (\dot x x')^2 } dtdz
\ee
can be expressed via $k_R^{\m}$ and $k_L^{\m}$.

To normalize the infinite string  we have to perform a renormalization
putting      $ \alpha^{\prime -1}\int (x')^2 dz = m, $
which yields the usual action for the center of mass of a pointlike particle
\be
S= m\int \sqrt{( \dot x)^2 }dt.
\ee

For the system of two D-strings  in the rest one can use the gauge with
$\dot x^0 =1, \quad \dot x^a=0,$ where the term   $(\dot x x')^2$ drops out,
and  the action takes the form
\be
S=\alpha^{\prime -1}\int dt \int \sqrt{p^a p_a} d\sigma,
\ee
where
\be
p^a = \d _\sigma x^a = \frac 12 [ x_R^{\prime\m}(t+\sigma) -
x_L^{\prime\m}(t-\sigma)].
\ee

However, one of the most important arguments in favor of the
combination of two chiral strings is suggested by analogue to
the Dirac equation, which has to be obtained for the Kerr spinning
particle if it has a relation to the structure of electron.
It is known that in the Weyl basis the Dirac current can be
represented as a sum of two lightlike components of opposite chirality
\be
J_\m = e (\bar \Psi \gamma _\m \Psi) = e (\chi ^{+} \sigma _\m  \chi +
\phi ^{+} \bar \sigma ^\m  \phi ),
\ee
where
\begin{equation}
\Psi =
\left(\begin{array}{c}
 \phi _\alpha \\
\chi ^{\dot \alpha}
\end{array} \right),
\label{Psi}\end{equation}
and
\be
\bar\Psi =(\chi ^+,
\phi ^+ )
\label{barPsi}
\ee
Two real lightlike 4-vectors $k_L^{\m},\quad k_R^{\m}$
can be expressed in spinor form
\be
k_L^{\m} =  \phi ^{+} \bar\sigma ^\m  \phi
\qquad k_R^{\m} =  \chi ^{+} \sigma ^\m  \chi,
\ee
and two extra complex null vectors can be formed
\be
m^{\m} =  \chi ^{+} \sigma ^\m  \phi
\qquad \bar m^{\m} =  \phi ^{+} \sigma ^\m  \chi
\ee
which complete the null tetrad. This  analogue shows that the Dirac
equation can only describe the axial stringy system of the Kerr spinning
particle.

Indeed, the solution
containing combination of three terms with $n=-1, 0, 1$ represents  also
especial interest since it yields
a smooth e.m. field packed along the Kerr string with
one half of the wavelength and gives an electric charge to the solution.

Note, that orientifold structure of the Kerr circular string
admits apparently the excitations with $n= \pm 1/2$ too, so far as the
negative half-wave can be packed on the covering space turning into
positive one on the second sheet of the orientifold. However, the meaning
of this case is unclear yet, and it demands a special consideration.

\section{Einstein-Maxwell axial pp-wave solutions}

The e.m. field given by (\ref{cFLR}), (\ref{fR}) and (\ref{fL}) can be
obtained from the potential

\be
\cA= - AZ e^3 - \chi d \Y ,
\label{cA}
\ee
where $A=\psi /P^2$ is given by (\ref{A}) and
\be
\chi = \int P^{-2} \psi d Y,
\label{chi}
\ee
$\Y$ being kept constant in this integration.
The considered wave excitations have the origin from the term
\be
\cA= P^{-2}\psi _n Z e^3 = q Y^n \exp {i\omega \t} P^{-2} Z e^3
\ee
and acquire the following asymptotical $z^{\pm}$ forms:

For $n=1; z<0$

\be
\cA ^- =q Y e^{i\omega \t}(r+ia \cos \theta)^{-1} e^3/P \simeq
-2 q \frac { e^{i\omega _1 \t_- +i\phi}}{\rho} dv
\ee

For $n=-1; z>0$

\be
\cA ^+ =q Y^{-1} e^{i\omega \t}(r+ia \cos \theta)^{-1} e^3/P \simeq
2 q \frac { e^{i\omega _{-1}\t_+ -i\phi}}{\rho} du;
\label{Au}
\ee

\begin{figure}[ht]
\centerline{\epsfig{figure=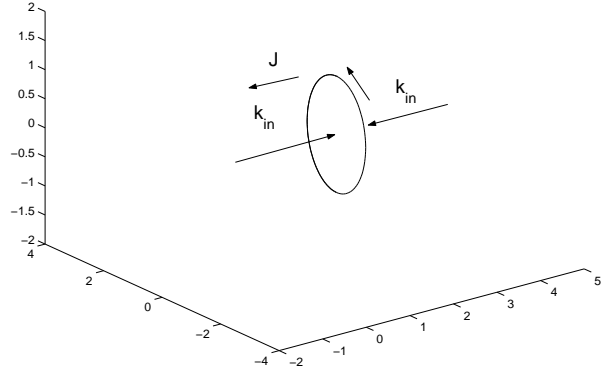,height=6cm,width=8cm}}
\caption{Schematic description of the Kerr antiparticle.
Two axial singular strings are directed ``inward''.}
\end{figure}

\begin{figure}[ht]
\centerline{\epsfig{figure=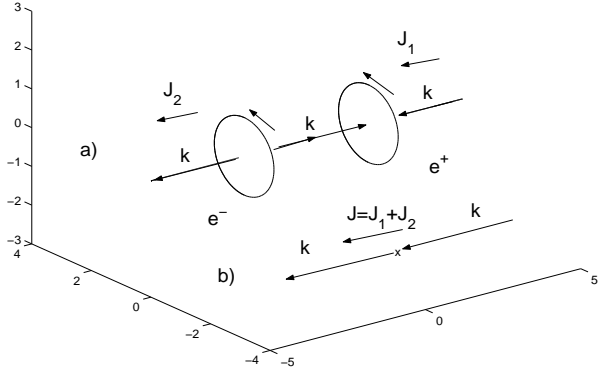,height=6cm,width=8cm}}
\caption{ (a) annihilation of the Kerr particle and antiparticle and
(b) formation of the lightlike particle.}
\end{figure}

Each of the partial solutions
represents the singular plane-fronted e.m. wave propagating along
$z^+$ or $z^-$ half-axis without damping.
It is easy to point out the corresponding
self-consistent solution of the Einstein-Maxwell field equations which
belongs to the well known class of pp-waves \cite{KraSte,Per}.

The metric has the Kerr-Schild form
\be g_{\m\n}
= \h_{\m\n} + 2 h k_{\m} k_{\n}, \label{ks} \ee
where function $h$ determines the Ricci tensor
\be
R^\mn = - k^\m k^\n \Box h,
\label{Rmn}
\ee
 $k^\m= e^{3\m}/P$ is the normalized principal null direction
(in particular, for the $z^+$ axis $k^\m dx^\m= - 2^{1/2}du $), and
$\Box$  is a flat D'Alembertian
\be
\Box=2\d _\z \d _\Z + 2 \d _u \d _v  \ .
\label{Box}
\ee
The Maxwell equations take the form
\be
\Box \cA = J=0
\label{BoxA}
\ee
and can easily be integrated leading to the solutions
\bea
 \cA ^+ = [ \Phi ^+(\z) + \Phi ^-(\Z) ] f^+(u,v) du, \\
 \cA ^- = [ \Phi ^+(\z) + \Phi ^-(\Z) ] f^-(u,v) dv,
\label{cApm}
\eea
where $\Phi ^{\pm}$ are arbitrary analytic functions, and functions
$f^\pm $
describe the arbitrary retarded and advanced waves.
In our case we have the retarded-time parameter
$\t = t - r - ia \cos \theta$ which takes at the $z^+$
axis the values $\t \simeq - 2^{1/2} u -ia $ and at the $z^-$
axis the values $\t \simeq 2^{1/2} v + ia $. Therefore, we have
\be
f^+ = f^+(u), \quad f^- = f^-(v).
\label{fpm}
\ee
The corresponding energy-momentum tensor  will be
\be
T^\mn = \frac 1 {8\pi} |\cF^+_{-1}|^2 k^\m k^n ,
\label{Tmn}
\ee
where for $z^+$ wave $k_\m dx^\m = -2^{1/2}du $ and
for $z^-$ wave $k_\m dx^\m = 2^{1/2}dv $.

The Einstein equations $R^\mn = -8 \pi T^\mn$ take the simple
asymptotic form
\be
\Box h = |\cF^+_{-1}|^2 = 16 q^2e^{-2a\omega}\rho^{-4}.
\label{heq}
\ee
This equation can  easily be integrated and yields the singular solution
\be h=  8 q^2e^{-2a\omega}\rho^{-2} .
\label{h}
\ee
Therefore, the wave excitations of the Kerr ring lead to the appearance
of singular pp-waves which propagate outward along the $z^+$ and/or
$z^-$ half-axis.

These axial singularities are evidences
of the axial stringy currents, which are exhibited explicitly
when we try to regularize the singularities \cite{BurWit} on the base of the
Witten field model for the cosmic superconducting strings \cite{Wit}.

The resulting excitations have the Compton wave length which is
determined by the size of the Kerr circular string.
However, for the moving systems the excitations of the axial stringy system
are modulated by de Broglie periodicity.

One can see here a striking similarity  with the well known
elements and methods of the signal transmission in the systems of radio
engineering and in the radar systems. In fact, the chiral axial string
resembles a typical system for the signal transmission containing
a carrier frequency which is modulated by the signal - the carrier of
an information. One sees that the Kerr circular string can also
be considered as a generator of the carrier frequency, and
plays a role of the antenna. Basing on the principle that the fine
description of a quantum system has to absorb  maximally the known classical
information on this system, one can conjecture that the above strikingly
simple structure can have a relation to the structure of spinning particles.

In conclusion, one can conjecture which changes could correspond to the
 Kerr anti-particle. It has to be the change of the PNC direction,
as it is shown in fig.4. It yields a natural picture of annihilation
as it is shown in the fig.5. It was discussed in \cite{BurMag} that
the size of the Kerr circular string for the massless Kerr spinning
particle has to grow to infinity and disappear. As a result there retains
only a single chiral string.

\section*{Acknowledgments}
The most part of this work was reported at the XXVI Workshop on the Fundamental
problems of HEP and Field Theory (IHEP, Protvino, July 2003)
and at the Vigier IV Conference (Paris, September 2003), and we
 are thankful to Organizers of these meetings for very kind invitation and
financial support. This work was also partially supported from the International
Science Education Project and we are very thankful to Jack Sarfatti for this
support and very useful discussions.

\section*{Appendix}
The leading wave terms for $|n|\le 1$ are the following:

\be
\cF^-_1=\frac {4q e^{i2\phi +i\omega _1 \t _-}}
{\rho ^2} \ d \Z \wedge dv, \qquad
\cF^+_{-1}= -
\frac {4q e^{-i2\phi+i\omega _{-1} \t _+}} {\rho ^2} \ d \z \wedge du ,
\ee

{\bf For $n=1$.}

At $z^-$ half-axis

\be
\cF^-_1=4 q \frac {e^{i2\phi}+i\omega _1 \t _- } {\rho ^2} \ d \Z \wedge dv -
\frac {q e^{i\omega _1 \t _-}} {r ^2} \ d \z \wedge du,
\ee

and at $z^+$ half-axis

\be
\cF^+_1=\frac {3q e^{i2\phi + i\omega _1 \t _+} \sin ^2 \theta } {4 r ^2} \
d \Z \wedge dv +
\frac {q e^{i\omega _1 \t _+ }} {r ^2} \ d \z \wedge du .
\ee

{\bf For $n=0$.}

At $z^-$ half-axis

\be
\cF^-_0=q \frac {(1+2a\omega) e^{i\phi +i\omega _0 \t_- } \sin \theta } {r ^2} \
 d \Z \wedge dv -
q\frac {e^{-i\phi+i\omega _0 \t _-}\sin \theta} {r ^2} \ d \z \wedge du,
\ee

and at $z^+$ half-axis

\be
\cF^+_0=
q\frac {e^{i\phi+ i\omega _0 \t_+ } \sin \theta } {r ^2} \ d \Z \wedge dv
+q\frac {(2a\omega -1) e^{-i\phi + i\omega _0 \t_+  }\sin \theta} {r ^2}
 \ d \z \wedge du.
\ee

{\bf For $n=-1$.}

At $z^-$ half-axis

\be
\cF^-_{-1}=-\frac {q e^{i\omega _{-1} \t_-  }} {r^2} \ d \Z \wedge dv -
\frac {3q e^{-i2\phi+i\omega _{-1} \t_-  } \sin ^2 \theta }
{4 r ^2} \ d \z \wedge du.
\ee

and at $z^+$ half-axis

\be
\cF^+_{-1}=\frac {q e^{ i\omega _{-1} \t_+ } } {r ^2} \ d \Z \wedge dv -
\frac {4q e^{-i2\phi +i\omega _{-1} \t_+  }} {\rho ^2} \ d \z \wedge du .
\ee

{\bf Similar, for $n=1/2$.}

At $z^-$ half-axis

\be
\cF^-_{1/2}=q\frac {2^{1/2} e^{i 3\phi /2}+i\omega _{1/2} \t_-  } {\rho ^{3/2} r^{1/2}}
 \ d \Z \wedge dv -
\frac {3q  e^{-i\phi/2+i\omega _{1/2} \t_-  } \sin ^{1/2} \theta} {2^{3/2} r ^2} \
d \z \wedge du,
\ee

and at $z^+$ half-axis

\be
\cF^+_{1/2}=
\frac {5q e^{i3\phi/2+i\omega _{1/2} \t_+ } \sin ^{3/2}\theta } { 2^{1/2} \ 4 \ r ^2}
 \ d \Z \wedge dv +
\frac {qe^{-i\phi / 2 +i\omega _{1/2} \t_+}} {2^{1/2}\rho ^{1/2} r ^{3/2}}
 \ d \z \wedge du .
\ee

{\bf For $n=-1/2$.}

At $z^-$ half-axis

\be
\cF^-_{-1/2}=\frac {-q e^{i\phi /2+i\omega _{-1/2} \t_- }} {2^{1/2}\rho ^{1/2} r^{3/2}}
 \ d \Z \wedge dv -
\frac {5q e^{-i3\phi/2+i\omega _{-1/2} \t_-  } \sin ^{3/2}  }
{ 2^{1/2} \ 4 \ r ^2} \ d \z \wedge du,
\ee

and at $z^+$ half-axis

\be
\cF^+_{-1/2}=
\frac {3q e^{i\phi /2 +i\omega _{-1/2} \t_+  } \sin^{1/2} \theta} {2^{3/2} r^{2}} \
d \Z \wedge dv
-\frac {2^{1/2}q e^{-i3\phi /2 +i\omega _{-1/2} \t_+ }} {\rho ^{3/2} r ^{1/2}}
 \ d \z \wedge du.
\ee

\end{document}